\documentclass[12pt,preprint]{emulateapj}
\usepackage{graphicx}
\usepackage{subfigure} 
\usepackage{natbib} 
\usepackage{float}
\usepackage{apjfonts}
\def\swift{{\it Swift}}
\def\chandra{{\it Chandra}}
\usepackage{color}

\shorttitle{SN to SNR Transition: SN~2013by} 
\shortauthors{Black et al.}

\begin{document}

\title{The Transition of a Type IIL Supernova into a Supernova Remnant:\\
Late-time Observations of SN\,2013by}

\author{C. S.\ Black\altaffilmark{1}, D. Milisavljevic\altaffilmark{2,3}, R. Margutti\altaffilmark{4}, R. A. Fesen\altaffilmark{1},  D.\ Patnaude\altaffilmark{2}, S.\ Parker\altaffilmark{5}} 
\affil{\altaffilmark{1}6127 Wilder Lab, Department of Physics \& Astronomy, Dartmouth College, Hanover, NH 03755 }
\affil{\altaffilmark{2}Harvard-Smithsonian Center for Astrophysics, 60 Garden St., Cambridge, MA 02138}
\affil{\altaffilmark{3}Department of Physics and Astronomy, Purdue University, 525 Northwestern Avenue, West Lafayette, IN 47907, USA}
\affil{\altaffilmark{4}Center for Interdisciplinary Exploration and Research in Astrophysics (CIERA) and Department of Physics and Astronomy, Northwestern University, Evanston, IL 60208}
\affil{\altaffilmark{5}Parkdale Observatory, 225 Warren Road, RDl Oxford, Canterbury 7495, New Zealand}

\begin{abstract}

We present early-time \swift\ and \chandra\ X-ray data along with late-time optical and near-infrared observations of SN~2013by, a Type IIL supernova (SN) that occurred in the nearby spiral galaxy ESO 138$-$G10 ($D \sim 14.8$ Mpc). Optical and NIR photometry and spectroscopy follow the late-time evolution of the supernova from days +89 to +457 post-maximum brightness. The optical spectra and X-ray light curves are consistent with the picture of a SN having prolonged interaction with circumstellar material (CSM) that accelerates the transition from supernova to supernova remnant (SNR).  Specifically, we find SN~2013by's H$\alpha$ profile exhibits significant broadening ($\sim$ 10,000 km\,s$^{-1}$) on day +457, the likely consequence of high-velocity, H-rich material being excited by a reverse shock. A relatively flat X-ray light curve is observed that cannot be modeled using inverse-Compton scattering processes alone but requires an additional energy source most likely originating from the SN-CSM interaction. In addition, we see the first overtone of CO emission near 2.3 $\micron$ on day +152, signaling the formation of molecules and dust in the SN ejecta and is the first time CO has been detected in a Type IIL supernova. We compare SN~2013by to Type IIP supernovae whose spectra show the rarely observed SN-to-SNR transition in varying degrees and conclude that Type IIL SNe may enter the remnant phase at earlier epochs than their Type IIP counterparts.

\end{abstract}


\section{Introduction}

Type II supernovae, the most common type of core-collapse stellar explosion, are divided into two broad classes, Type IIL and IIP, based on their lightcurves (CCSNe; \citealt{Li11}). They are associated with red supergiant progenitor stars that have initial masses between 8-20 M$_{\odot}$, but the origins of their observational diversity are not fully understood \citep{Sma09,Val16}.

Type IIP SNe (SN IIP) are defined by the nearly uniform plateaus in their lightcurves that last $\sim$100 days \citep{Arc12,Far14}.  In contrast, Type IIL (SN IIL) light curves decline linearly and exhibit a magnitude change of more than 0.5 mags in the first 50 days post maximum brightness (s50$_\textrm{v} > $ 0.5 mag) \citep{Li11,Far14}.  SNe IIP generally have fainter peak magnitudes $\approx -16.0$, while SNe IIL are the most luminous SNe II with a nearly uniform peak magnitude of roughly -17.5 \citep{Fil97,Li11}, though they can reach magnitudes of up to -19.3 at their brightest \citep{Ric02}.

The prevailing view is that the progenitors of Type IIL SNe are higher mass stars ($\la$20 M$_\odot$) that have larger radii and less hydrogen than the progenitors of most Type IIP  \citep{Blinnikov93,Fil97}.  SN IIL also produce more O-rich ejecta and typically have stronger [\ion{O}{1}]/H$\alpha$ ratios than the other Type II subclasses \citep{Far14}.

However, some recent studies suggest that Type IIP and IIL SNe may be closely related \citep{And14,San15,Val15}.  Late-time photometry of SNe~IIL can exhibit a significant drop in magnitude roughly 100 days past maximum brightness, resembling the one seen in Type IIP SN after the plateau phase.  This suggests that the nebular phases of these subtypes may be similar \citep{Val15}.

In addition, it is possible to recreate Type IIL and IIP lightcurves using the same progenitor but with varying degrees of circumstellar material (CSM) surrounding the star.  This may explain why some SN appear to take on characteristics of multiple Type II subclasses as they evolve, like SN~2013fs \citep{Mor17}.

\begin{figure*}[!ht]
	\centering
	\includegraphics[scale=0.8]{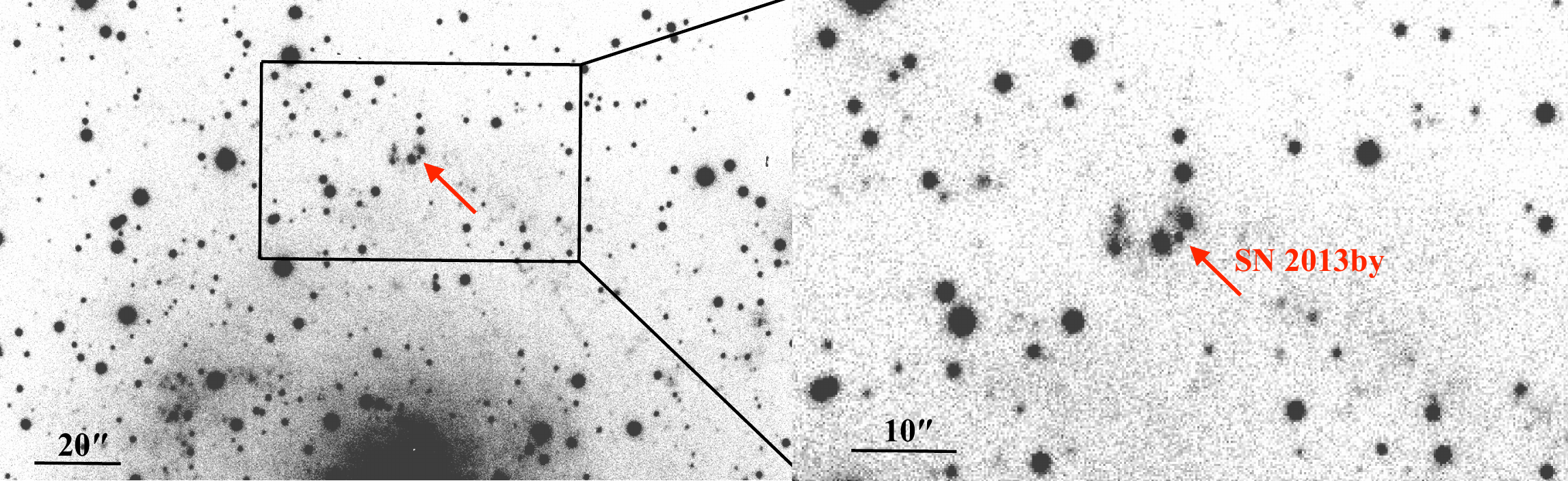}
	\caption{Sloan $r'$ image of SN~2013by in ESO 138-G10 from day +457 obtained with the 6.5m Magellan Baade telescope. The right panel shows the boxed region from the left panel.  North is up, East is to the left.}
	\label{fig:13by}
    \vspace{0.25cm}
\end{figure*}

SN~2013by was discovered on 23 April 2013 in the galaxy ESO 138-G10  at coordinates $\alpha$(2000) = 16$^{\rm h}$ 59$^{\rm m}$ 02.43$^{\rm s}$, $\delta$(2000) = $-$60$^\circ$ 11$^{\prime}$ 41${\farcs}$8 by the Backyard Observatory Supernova Search (BOSS)
\citep{Par13}.  \citet{Val15} observed it to have a short rise time of roughly 10 days, reaching a peak absolute magnitude of M$_V$ = -18.2 on 1 May 2013 assuming a distance of 14.8 Mpc. The SN then declined at a rate of s50$_\textrm{v}$ = 1.46 mag, placing it well above the SN IIL threshold defined by \citet{Li11}.  This was followed by a sharp decline in magnitude some 80 days post explosion.

Spectra obtained in the first few weeks post-shock breakout showed features consistent with moderately interacting SNe II/IIn \citep{Mil13}. A comprehensive investigation by \citet{Val15} led them to conclude that SN ejecta interacted with surrounding circumstellar material for more than one month after post-max.

Here we report on early X-ray observations from +2 to +70 days post-maximum brightness plus optical and near infra-red observations from +89 to +457 days of SN~2013by. In \S\ref{sec:Data} we present our data and observations, in \S\ref{sec:Results} we discuss the results of our photometry and spectroscopy, and in \S\ref{sec:Disc} we investigate the implications these results have on the SN environment and the progenitor system. Throughout this work we adopt the NASA Extragalactic Database\footnote{http://ned.ipac.caltech.edu/} distance to the host galaxy ESO 138-G10 to be $14.8\pm1$ Mpc ($\mu$ = $30.84 \pm 0.15$), which is derived from the radial velocity of $1144 \pm 2$ km\,s$^{-1}$ \citep{Kor04} assuming H$_0 = 73 \pm 5$ km s$^{−1}$ Mpc$^{-1}$ after correcting for Local Group infall towards the Virgo cluster.

\section{Data Set}\label{sec:Data}

\subsection{Optical and Near Infrared Observations}
In Figure \ref{fig:13by} we present Sloan $r'$ images of SN~2013by in ESO 138-G10 obtained with the 6.5m Magellan Baade telescope at the Las Campanas Observatory. At day +457 the SN can be seen between two adjacent sources north of the host galaxy's center. Multiple epochs of low-resolution optical spectra of SN~2013by were also obtained. 

The spectra were taken with the IMACS f/2 camera using the E2V CCD with a 300 lines mm$^{-1}$ 6700 \AA\ grating in combination with a 0$\farcs$7 wide slit.  A set of $2 \times 1200$ s exposures were taken on 29 September 2013 (day +151) and 3 February 2014 (day +278), and $2 \times 1800$ s exposures were obtained on 1 August 2014 (day +457).  Acquisition images in the Sloan $r'$ filter, which are listed in Table~\ref{tab:PhotData}, were obtained at each epoch and used for our photometry. 

Three epochs of  near-infrared spectra were also obtained in 2013 and 2014 at days +89, +152, and +279 with the Baade telescope using the FoldedPort Infrared Echellette (FIRE; \citealt{Sim08}) in combination with the low-dispersion prism, a $0\farcs 6$ wide slit, and a Hawaii-2RG detector. The spectral resolution was $R \sim 500$ in $J$-band. Total exposure times were between 1200-1800s and split up into individual 158s integrations that were dithered along the slit in an ABBA pattern. Details of all our observations can be found in Tables \ref{tab:PhotData} and \ref{tab:SpecData}.

Data reduction of the images and optical spectra
was done using \textsc{iraf}\footnote{\textsc{IRAF} is distributed by the National Optical
Astronomy Observatories, which are operated by the Association of Universities
for Research in Astronomy, Inc., under cooperative agreement with the National
Science Foundation.} and consisted of bias and background subtraction,
wavelength calibration, and aperture extraction.  FIRE data were reduced following standard procedures \citep{Hsi13} using a IDL pipeline (FIREHOSE). The host galaxy recession velocity was removed from all spectra. 

\begin{deluxetable}{lcccl}
  \tablecolumns{5}
  \tablewidth{\columnwidth}
  \tablecaption{Photometric Observations}
  \tablehead{
  \colhead{Date} & \colhead{Instrument} & \colhead{Exposure Time} & \colhead{Band} & \colhead{Mag}\\
  \colhead{ } & \colhead{ } & \colhead{(s)} & \colhead{} & \colhead{(r$'$)}}
  \startdata
  29 Sept 2013 & IMACS & 3 & Sloan r & 17.14 $\pm$ 0.07\\  
  3 Feb 2014 & IMACS & 30 & Sloan r & 17.99 $\pm$ 0.17\\
  1 Aug 2014 & IMACS & 10 & Sloan r& 20.78 $\pm$ 0.08
   \enddata
   \label{tab:PhotData}
\end{deluxetable}

\begin{deluxetable}{llcc}
  \tablecolumns{4}
  \tablewidth{\columnwidth}
  \tablecaption{Spectroscopic Observations}
  \tablehead{
  \colhead{Date} & \colhead{Instrument} & \colhead{Exposure Time} & \colhead{Wavelength} \\
  \colhead{ } & \colhead{ } & \colhead{(s)} & \colhead{Region}}
  \startdata
  29 July 2013 & FIRE & 8 x 158 & NIR\\
  29 Sept 2013 & IMACS & 2 x 1200 & Optical\\
  30 Sept 2013 & FIRE & 10 x 158 & NIR\\
  3 Feb 2014 & IMACS & 2 x 1200 & Optical\\
  4 Feb 2014 & FIRE & 10 x 158 & NIR\\
  1 Aug 2014 & IMACS & 2 x 1800 & Optical
  \enddata
  \label{tab:SpecData}
\end{deluxetable}

\begin{figure}
	\vskip -0.0 true cm
	\centering
    \subfigure[]{ \label{fig:XRT}
		\includegraphics[width=0.75\columnwidth]{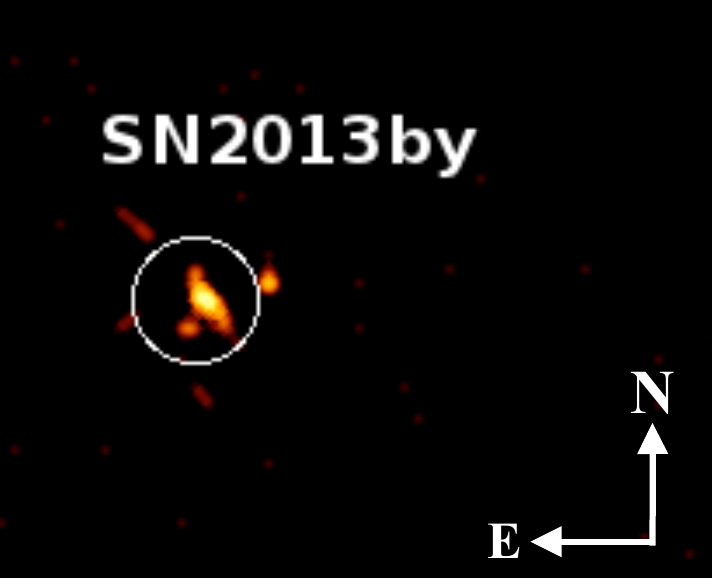}}
    \subfigure[]{ \label{fig:Chandra}
    	\includegraphics[width=0.75\columnwidth]{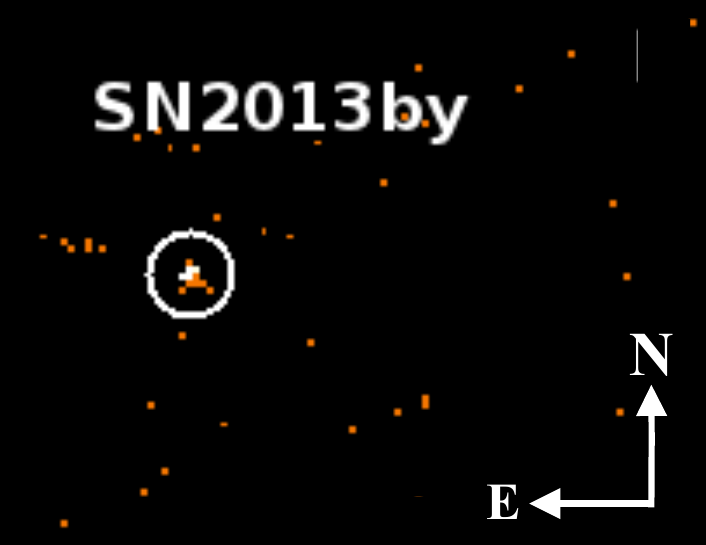}}
	\caption{(a) X-ray emission from SN\,2013by as detected by the Swift-XRT (0.3-10 keV). The image is comprised of observations collected starting April 24, 2013 until June 13, 2014, total of $16.5$ ks. White circle: 10" region centered at the SN position. (b) X-ray emission from SN\,2013by as detected by the Chandra X-ray Observatory (CXO) $\sim70$ days after the estimated shock breakout date. White circle: 3" region centered at the SN position.}
	\label{fig:Xray}
\end{figure}

\begin{figure}
	\centering
	\includegraphics[width=\columnwidth]{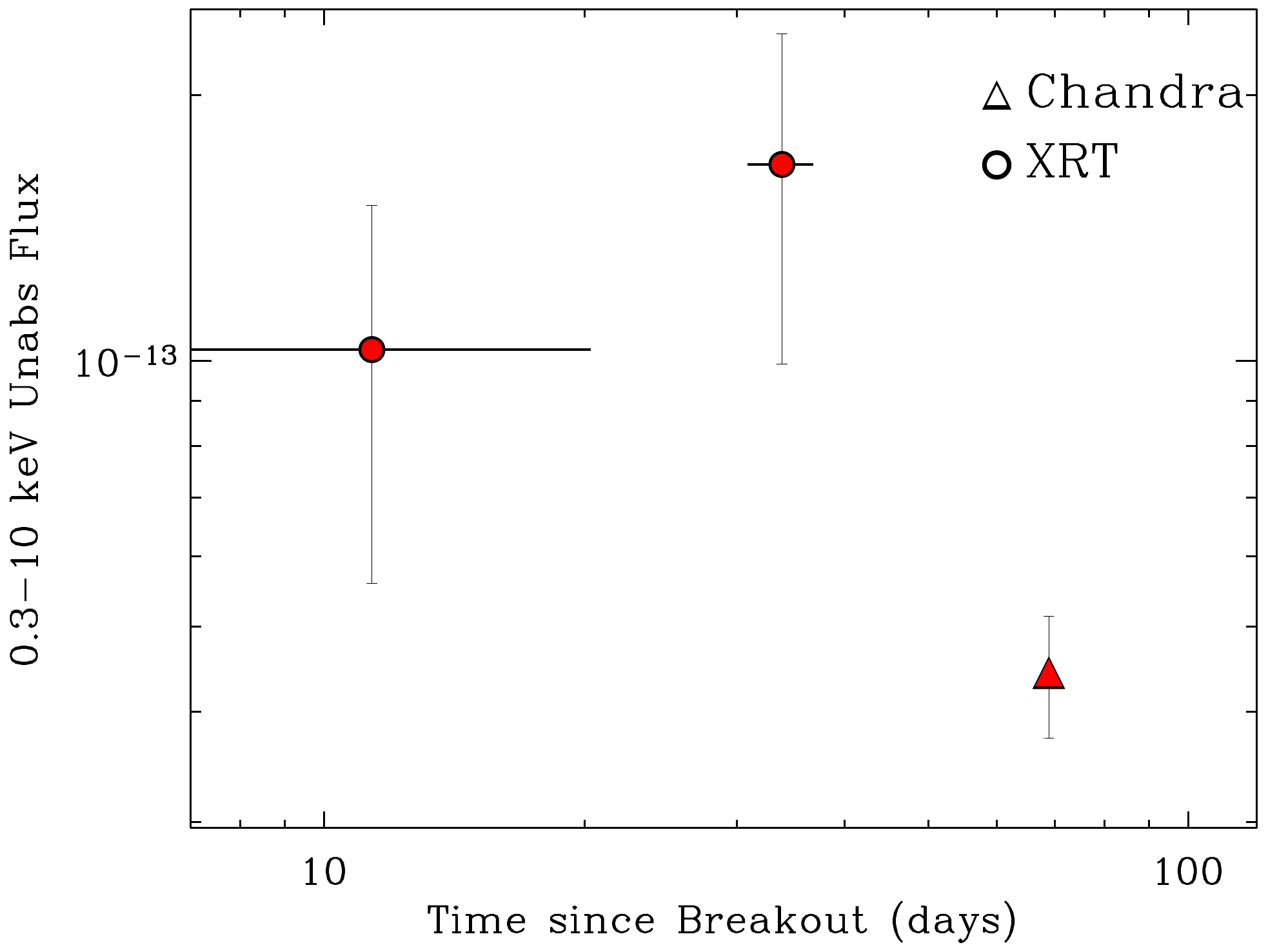}
	\caption{X-ray emission from SN\,2013by as detected by the \emph{Swift}-XRT (circles) and Chandra (triangle).}
	\label{fig:LC}
\end{figure}

\bigskip
\subsection{X-Ray Observations}
\label{Sec:XRay}

The \emph{Swift} X-Ray Telescope \citep{Geh04,Bur05}
started observing SN\,2013by on 24 April 2013, just 2 days after the estimated shock breakout \citep{Val15} (PI Margutti). XRT data was analyzed using  HEASOFT (v6.15) along with the corresponding calibration files.
Standard filtering and screening criteria was applied.  An X-ray source was clearly detected by the XRT at the position of SN\,2013by (Fig. \ref{fig:XRT}) until $\sim40$ days since breakout. 
At t $>40$ days the X-ray source was too faint and no longer detected by \emph{Swift}-XRT (Fig. \ref{fig:LC}).

Follow-up X-ray imaging of SN\,2013by by the \emph{Chandra} X-ray Observatory (CXO) with ACIS-S was initiated on 30 June 2013, corresponding to $\sim 70$ days since breakout (PI Pooley), see Figure \ref{fig:Chandra}. Data have been reduced with the CIAO software package (version 4.6) and corresponding calibration files.  Standard ACIS data filtering has been applied.

In the 9.8 ks \chandra\ observation, X-ray emission at the location of SN\,2013by is seen, with a significance of $15.5\,\sigma$. The spectrum is well modeled by an absorbed power-law with a spectral photon index $\Gamma=2.0\pm0.3$ ($1\sigma$) and Galactic neutral hydrogen absorption N$_{HI}=1.4\times10^{21}\,\rm{cm^{-2}}$ \citep{Kal05}. The corresponding unabsorbed 0.3-10 keV flux is $(4.4\pm0.7)\times10^{-14}\,\rm{erg\,s^{-1}cm^{-2}}$. We use the best-fitting spectral parameters above to flux to calibrate the entire X-ray data set.
The temporal evolution of the X-ray emission from SN\,2013by is presented in Figure \ref{fig:LC}.

\begin{figure}
        \centering
        \includegraphics [width=\linewidth]{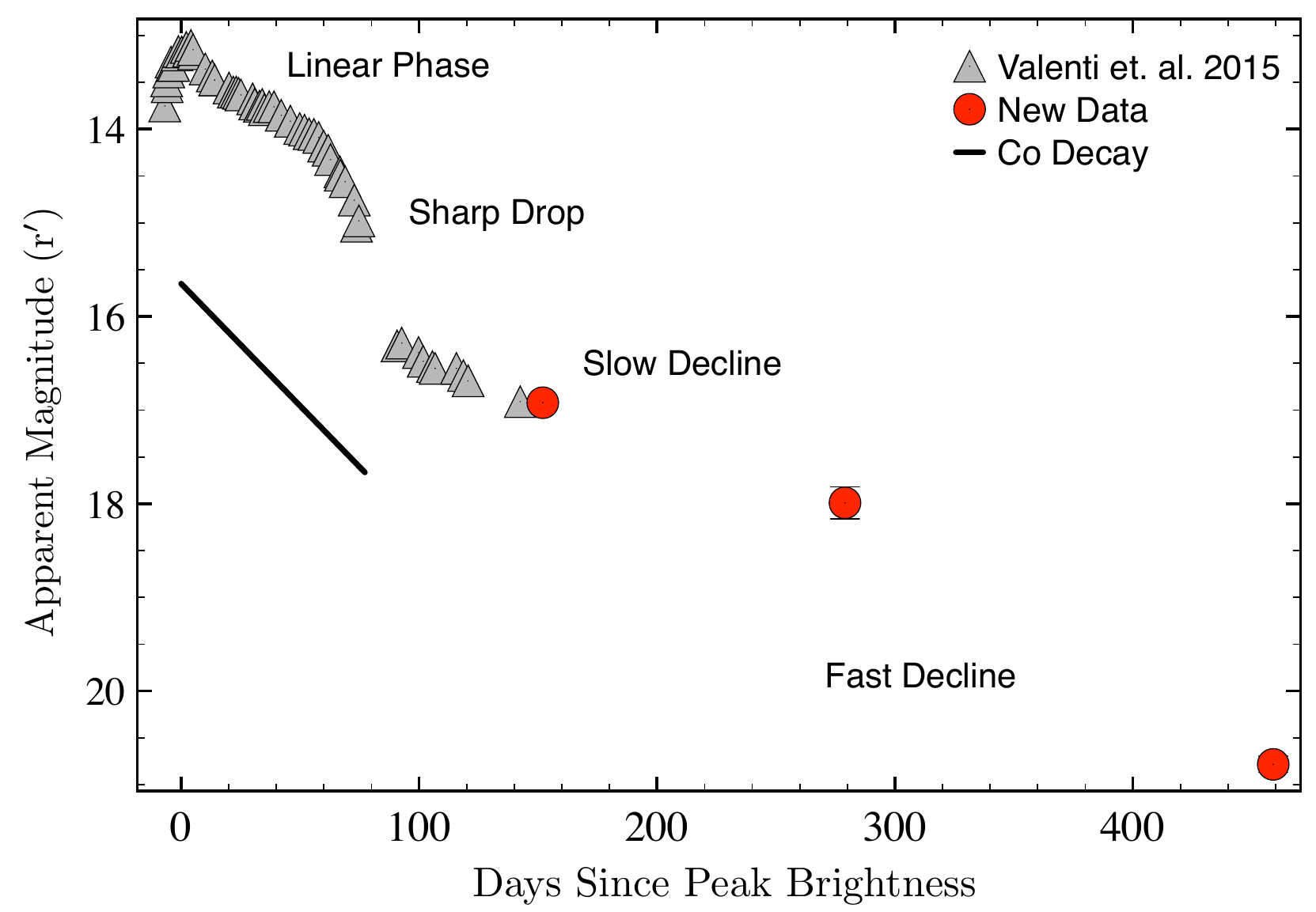}
        \caption{Sloan $r'$ photometry of SN\,2013by.  The gray triangles are data from \citet{Val15}, the red circles are the late-time data presented here, and the solid line represents the slope of heating by $^{56}$Co decay.}
        \label{fig:byPhot}
\end{figure}

\begin{figure*}
        \centering
		\includegraphics[width=\linewidth]{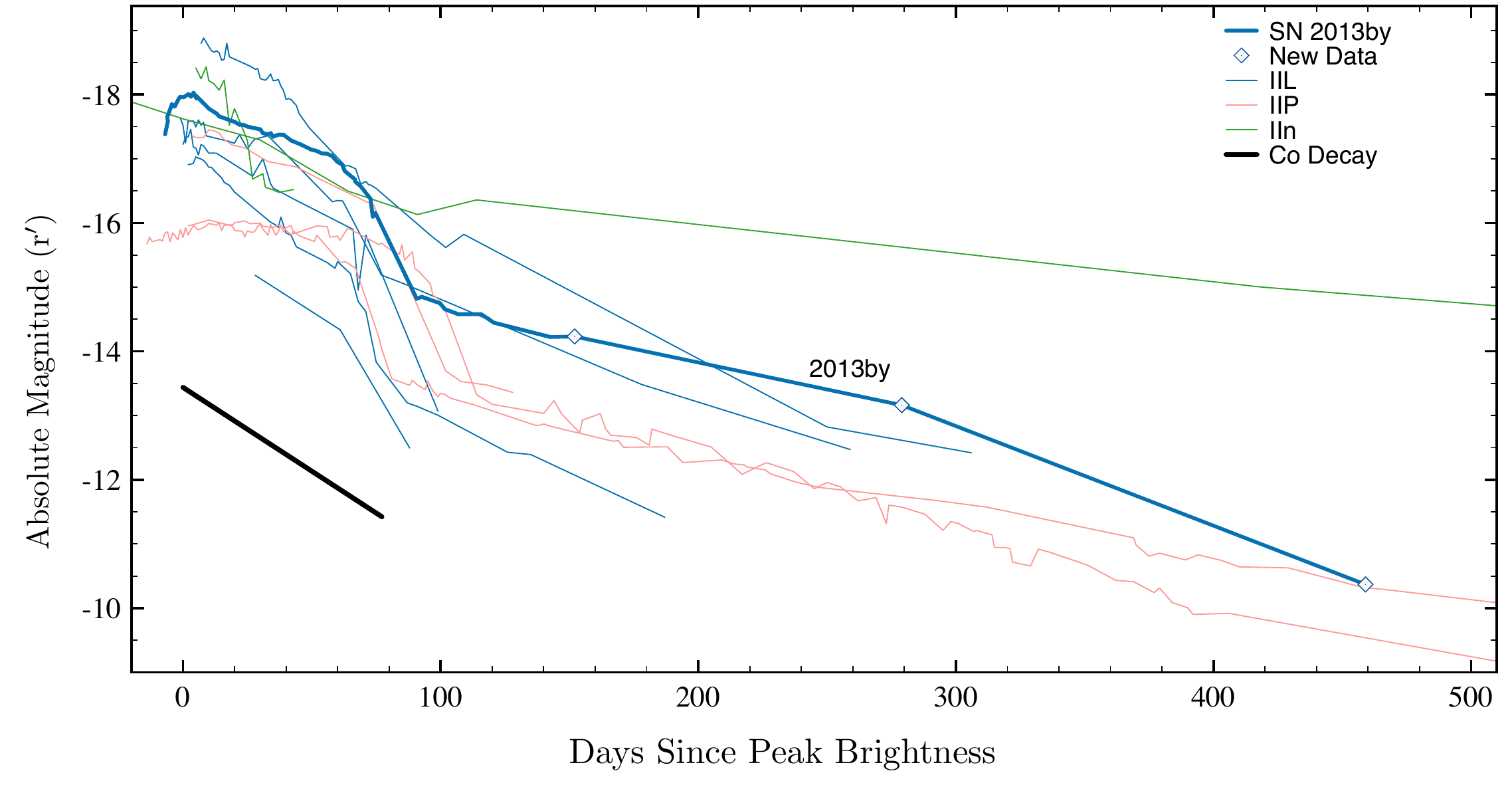}
        \caption{Comparing the lightcurves of Type IIP, IIL, and IIn SN using the Sloan r$'$ filter.  The bold blue line represents the photometry of SN~2013by and the diamonds are the new optical data presented here.  The thin blue lines correspond to SN IIL, the red depict SN IIP, the green lines represent SN IIn, and the black line shows the rate of Co decay.}
        \label{fig:IIPhot}
\end{figure*}

\begin{deluxetable}{lll}
  \tablecolumns{3}
  \tablecaption{Type II SNe shown in Figure \ref{fig:IIPhot}}
  \tablehead{
  \colhead{SN} & \colhead{Type} & \colhead{Reference}}
  \startdata
  1959D & IIL & \citealt{Arp61}\\
  1966B & IIL & \citealt{Gat67,Tsv83}\\
  1970G & IIL & \citealt{Bar73,Win74}\\
  1979C & IIL & \citealt{Vau81,Bar82}\\
  1980K & IIL & \citealt{BarCiaRos82,But82,Tsv83}\\
  1988Z & IIn & \citealt{Sta89,Pol89}\\
   & & \citealt{Gas89,Tur93}\\
   & & \citealt{Are99}\\
  2004dj & IIP & \citealt{Zha06}\\
  2004et & IIP & \citealt{Li05,Sah06}\\
   & & \citealt{Mag10,FarPoz14}\\
  2006jd & IIn & \citealt{Bro14}\\
  2008es & IIP & \citealt{MilCho09,Bro14}\\
  2010al & IIn & \citealt{Bro14}\\
  2011hw & IIn & \citealt{Smi12,Bro14}\\
  2013by & IIL & \citealt{Val15}\\
  2013ej & IIP & \citealt{Bro14}\\
  2014G & IIL & \citealt{Bro14}
   \enddata
   \label{tab:allSN} 
\end{deluxetable}



\begin{figure*}
        \centering
        \subfigure[]{ \label{fig:Optical_all}
		\includegraphics[width=0.85\linewidth]{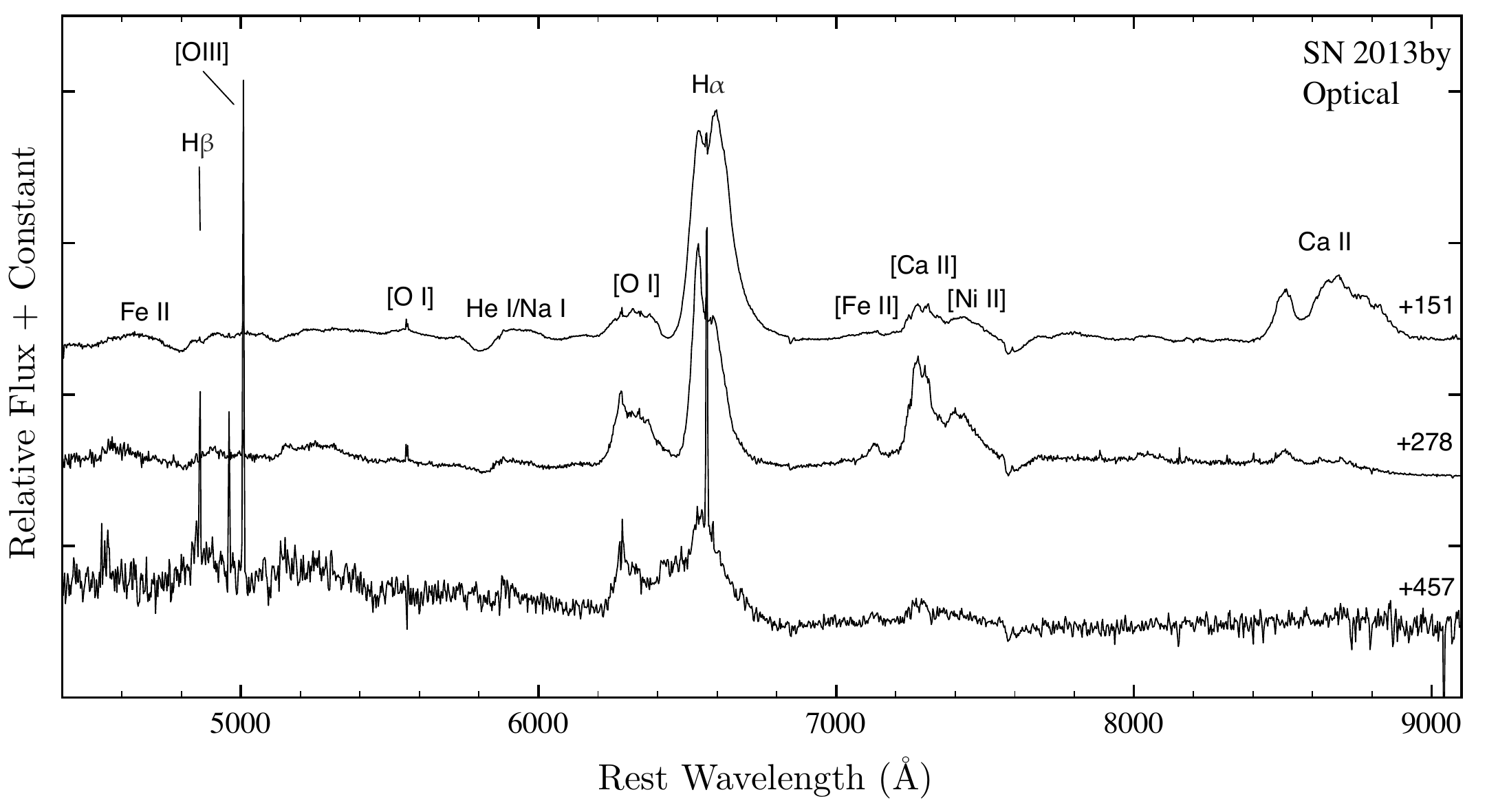} }
        \subfigure[]{ \label{fig:NIR_all}
		\includegraphics[width=\linewidth] {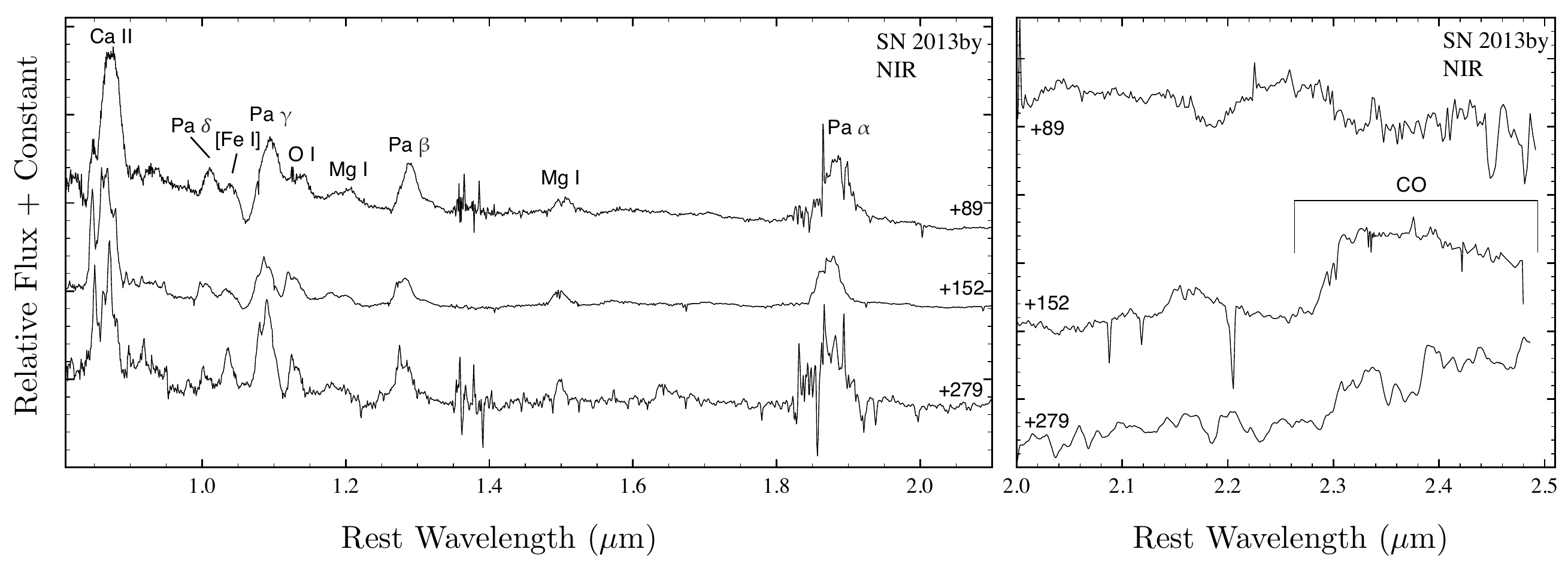} }
        \caption{(a) Optical spectra of SN~2013by at +151, +278, and +457 days post-max. Note the evolution of the asymmetry of the H$\alpha$ feature and the broadening of the blue wing at day +457. The strong, narrow [\ion{O}{3}] 4959,5007, H$\alpha$, and H$\beta$ lines seen in the optical spectra at day +457 may have significant contribution from a coincident, line-of-sight \ion{H}{2} region, and the increase in blue continuum flux is likely associated with nearby stars observed in our images (see Fig.~\ref{fig:13by}). (b) Bottom Left: NIR spectra of SN~2013by at +89, +152, and +279 days post-max. The Pa lines do not show evidence of asymmetry like that seen in H$\alpha$. Bottom Right: The NIR spectra has been scaled to show the CO emission.  The day +279 spectrum has been smoothed and is too noisy to confirm a CO detection.}
        \label{fig:Spec}
\end{figure*}

\begin{figure}
    \centering
	\includegraphics[width=\columnwidth]{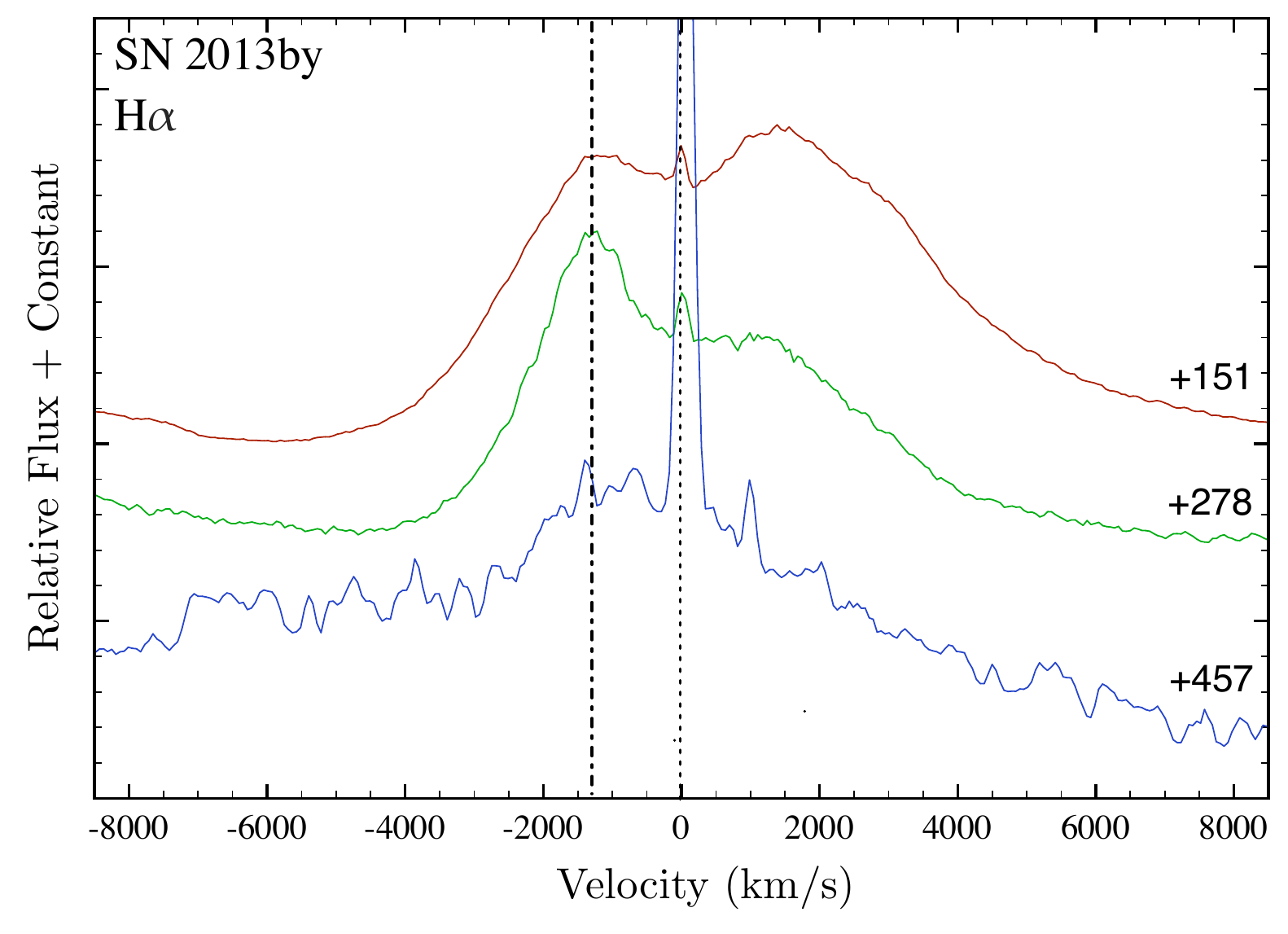}
    \caption{Emission line profiles of H$\alpha$ at days +151, +278, and +457.  The dotted line marks zero velocity and the dashed-dotted line marks the blueshifted components.}
    \label{fig:Ha}
\end{figure}


\section{Results}\label{sec:Results}

\subsection{Optical Photometry}\label{sec:Optical}

A plot of our Sloan $r'$ photometry of SN~2013by is shown in Figure \ref{fig:byPhot}. Gray triangles are photometry taken from \citet{Val15}, with the red circles indicating the new photometric observations presented here.  Together these data span from day -8 to +457 of the light curve evolution. 

The observed late-time slope is less steep than that expected solely from heating by radioactive $^{56}$Co decay (the solid line), where the SN initially fades at a rate of s50$_\textrm{v}$ = 1.46 mag.  This suggests that an additional energy source is powering its late-time luminosity.  SN~2013by's late-time light curve appears to decay at a fairly constant rate of roughly 0.45 mag per 50 days between days +100 and +278, but increases to around 0.77 mag per 50 days afterward. 

In Figure \ref{fig:IIPhot} we compare the Sloan $r'$ light curves of a sample of Type IIP, IIL, and IIn SNe out to epochs comparable to those presented for SN~2013by.  A list of the SNe used in this figure can be found in Table \ref{tab:allSN} and the $B$ and $V$ magnitudes were converted to Sloan $r'$ using prescriptions outlined in \citet{Jes05}.

The early light curve of SN~2013by decays at a rate similar to those observed for the few other SN IIL having published data.  However, after a steep drop around 80 days post-maximum light, the decay rate is similar to that seen in SNe IIP \citep{Val15}.

\subsection{Optical and NIR Spectroscopy}

Rest-frame optical and NIR spectra of SN~2013by are presented in Figure \ref{fig:Spec} with the strong emission features labeled. Significant evolution can be seen, most notably in the H$\alpha$ profile.  At day +151, H$\alpha$ is the dominant feature and shows an asymmetric triple peaked distribution where the emission is strongest at redshifted velocities, peaking at a velocity of around +1500 km/s.  On day +278, the red emission has diminished and the blue flux now dominates with a centroid near -1250 km\,s$^{-1}$. The overall shape and width of this feature remains roughly the same during this transition. 

However on day +457, the H$\alpha$ profile is no longer triple peaked and the ratio of H$\alpha$/[\ion{O}{1}] diminishes significantly. The blue wing of H$\alpha$ has increased from $-$4000 km~s$^{-1}$ on day +278 to $-$7500 km~s$^{-1}$ on day +457.  It is difficult to determine the full velocity extent of the blue wing of the H$\alpha$ profile due to its blending with the [\ion{O}{1}] 6300, 6364 lines.  The red wing has maintained a maximum velocity of roughly +4500 km~s$^{-1}$.  Figure \ref{fig:Ha} shows the emission line profiles of H$\alpha$ at all three epochs.

\begin{figure}
        \centering
        \includegraphics [width=\linewidth]{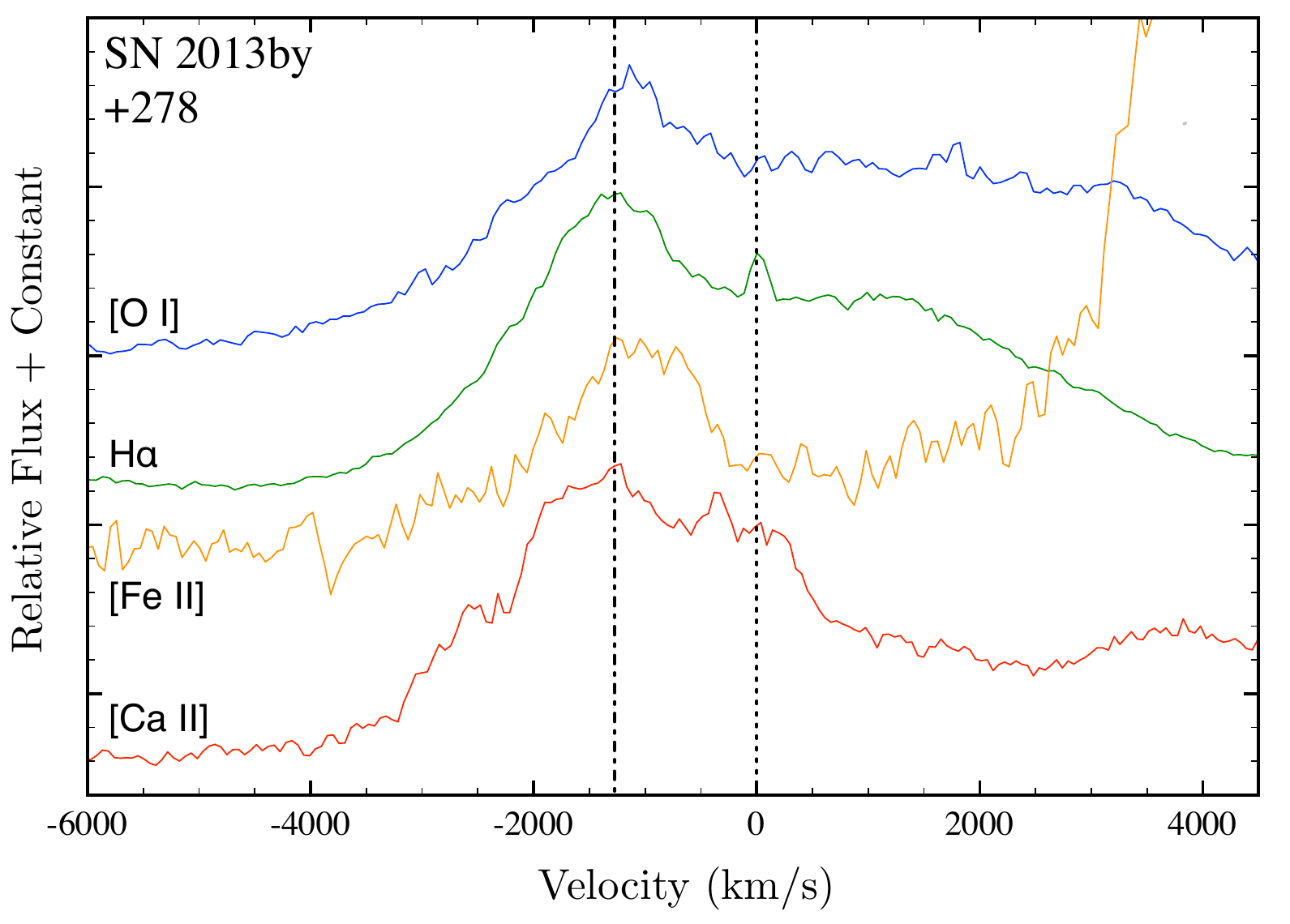}
        \caption{Emission line profiles of  [\ion{O}{1}] $\lambda\lambda$ 6300, 6364, H$\alpha$,  [\ion{Fe}{2}] $\lambda\lambda$7155, 7172, and [\ion{Ca}{2}] $\lambda\lambda$ 7291, 7324 at day +278, centered at 6300, 6563, 7155, and 7306 \AA\ respectively.}
        \label{fig:all_vel}
\end{figure}

The blueshifted peak observed in the day +278 H$\alpha$ profile is also seen in the [\ion{O}{1}], [\ion{Ca}{2}], and  [\ion{Fe}{2}] 7155 lines and Figure \ref{fig:all_vel} shows a comparison of these emission line profiles.  All four emission line profiles share the same blueshift of 1000 $-$ 1500 km/s. The emission peak is most prominent on day +278.  Many features fade significantly by day +457.

In the NIR (Fig.~\ref{fig:NIR_all}), the spectrum looks somewhat different and is dominated by Paschen emission features with the Pa$\alpha$ line extending from roughly $-$5000 km~s$^{-1}$ to $+$5500 km~s$^{-1}$. The NIR hydrogen lines do not show the same asymmetry as that seen in the optical line profiles, which suggests that the optical lines have been affected by optical depth effects and/or extinction due to internal dust.  The other prominent ions observed in the spectra include [\ion{Fe}{1}] 1.044 \micron, \ion{O}{1} 1.130 \micron, and \ion{Mg}{1} 1.503 \micron.  \ion{He}{1} 1.08 \micron\ is only distinctly seen at day +279. 

At day +152, emission from the first overtone of CO at 2.3 $-$ 2.5 $\micron$ can be seen in the right panel of Figure \ref{fig:NIR_all}.  This is the first time CO has been detected in a Type IIL SNe.  The +279 day spectrum lacks sufficient signal-to-noise to determine if CO is detected. CO formation is common in Type II SN and is typically seen +100 days after the SN explosion \citep{Che11,Yua16}.

The presence of CO signals the early formation of molecules and dust, as described by \citet{Ger00}.  This is consistent in the switch from a redshifted to a blueshifted asymmetry seen in the H$\alpha$ at day +278, where the emergence of CO and dust creation inhibit transmission of far side ejecta.

\begin{figure}
   \centering
   \includegraphics[width=\linewidth]{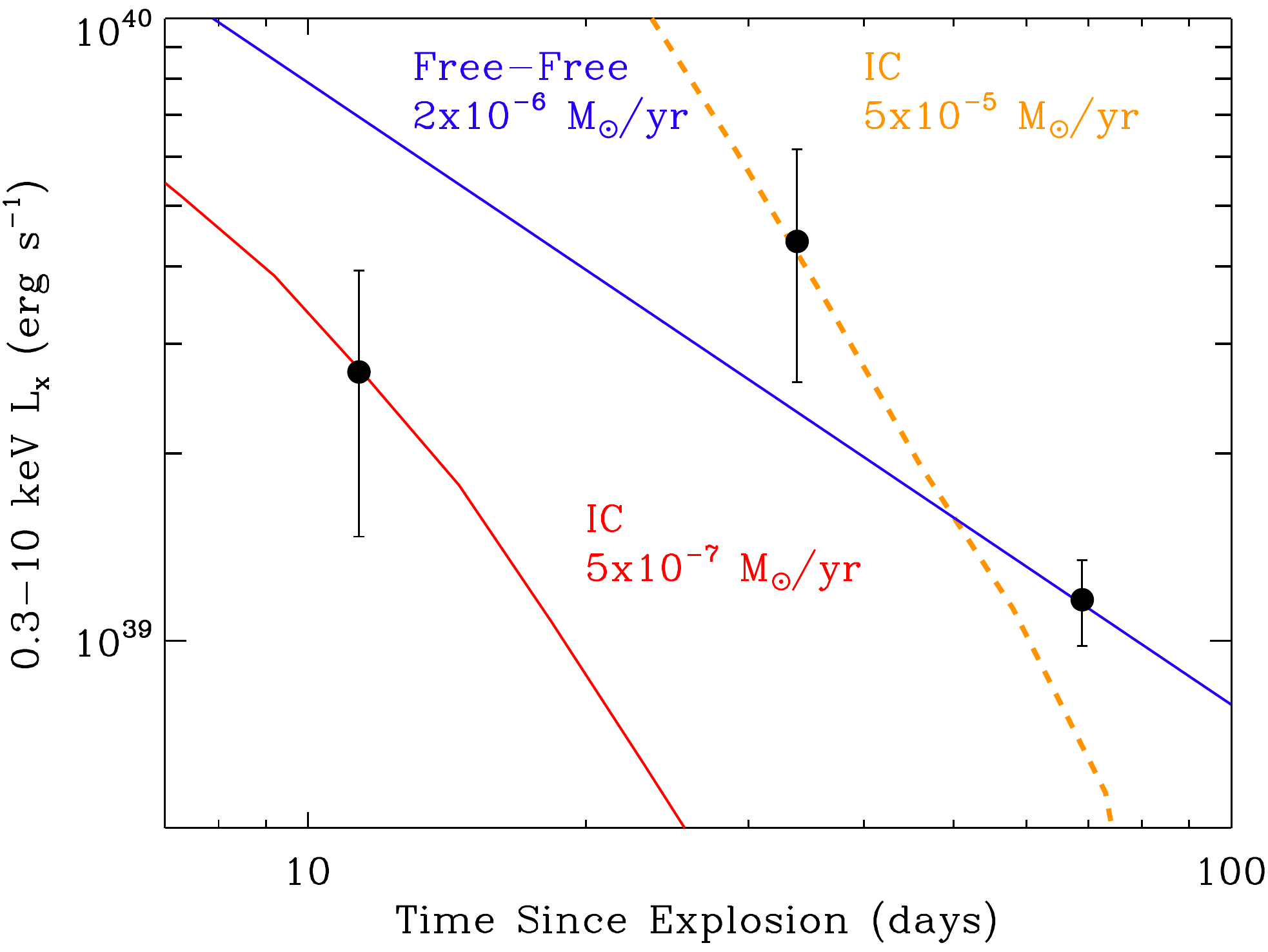}
   \caption{Models of X-ray emission that reasonably match our observations of SN\,2013byj require a combination of Inverse Compton scattering of optical photospheric photons by relativistic electrons at the SN shock and free-free thermal emission. The flat temporal evolution suggests that free-free emission dominates. We constrain the mass-loss rate of the progenitor star to be between  $\dot M\sim 10^{-4}\,\rm{M_{\sun}yr^{-1}}$ and $\dot M\sim (2-4)\times 10^{-6}\,\rm{M_{\sun}yr^{-1}}$ for a range of $n=10-15$, assuming v$_w$=10 km\,s$^{-1}$. }
   \label{fig:IC}
\end{figure}

\section{Discussion}\label{sec:Disc}

Our observations of SN\,2013by span late-time epochs rarely observed in SNe IIL. Together with data spanning days -8 to +109 previously reported in \citet{Val15}, our complementary analyses form a set of observations covering from -8 to +457 days post-maximum brightness.  The completeness of this data set enable us to follow the evolution of a Type IIL supernova and connect the properties of its early and late phases. 

\subsection{The SN-CSM Interaction}

Many features of SN\,2013by observed in the first 100 days post-explosion suggest that the SN interacted with local CSM. \citet{Mil13} reported that a narrow (FWHM $< 250$ km\,s$^{-1}$) H$\alpha$ profile observed in spectra obtained in the first week after discovery exhibited a blue-to-red asymmetry consistent with an intermediate Type IIL/IIn supernova.

\citet{Val15} reported high-velocity ($\sim$15,000 km~s$^{-1}$) H$\alpha$ absorption as well as a weak P-Cygni profile in their spectra obtained around this time and shortly afterward.  Both analyses drew the conclusion that SN\,2013by likely interacted with CSM that was less dense than material encountered by a typical Type IIn supernovae.

Our new observations at epochs beyond +151 days are consistent with this picture. The optical and X-ray light curves both suggest input of energy from SN-CSM interaction (Sections \ref{sec:Optical} and \ref{sec:Xray}). The triple-peaked H$\alpha$ emission line profile seen in our optical spectra at days +151 and +278 (Fig.~\ref{fig:Ha}) is not unlike that observed in SN\,1998S, which was interpreted as being due to the SN interacting with a toroidal distribution of clumpy CSM viewed edge on \citep{Ger00}. Interpreted in this way, the central unresolved (FWHM $< 100$ km~s$^{-1}$) component is the result of CSM that has been photoionized ahead of the forward shock (although contribution from coincident \ion{H}{2} region emission is possible), and the blue- and red-shifted broad components form as a result of interaction along the front and back segments of the disk.  

\begin{figure*}
        \centering
        \includegraphics [width=\linewidth]{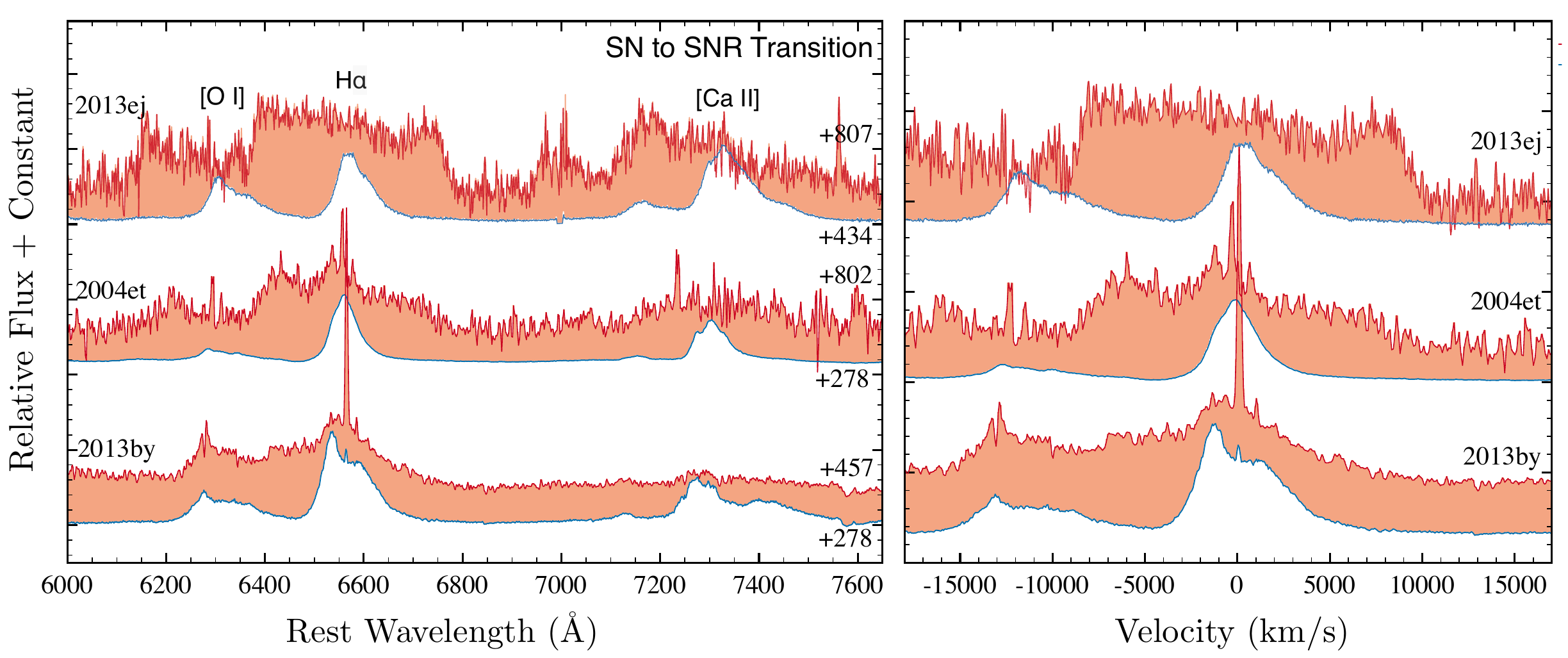}
        \caption{Late-time broadening of emission lines observed in SNe~2013ej, 2004et, and 2013by.  Blue spectra show epochs when SN emission is driven primarily by interior excitation by radioactive decay of Co$^{56}$.  The red spectra, which are shaded to highlight spectral evolution, exhibit significant broadening in H$\alpha$ and [\ion{O}{1}], suggesting that substantial ejecta at much higher velocities are being heated by the reverse shock.\vspace{0.6cm}}
        \label{fig:ReverseShock}
\end{figure*}

\subsection{X-ray Analysis}\label{sec:Xray}
X-ray emission in young SNe is associated with interaction between the forward blast wave and surrounding nearby ISM/CSM environment that has been  shaped by the stellar progenitor prior to core-collapse. Thus, our X-ray observations can be utilized to derive information about the mass-loss history of the progenitor star before stellar death \citep{Bjornsson04,Chevalier06}.  

Inverse Compton (IC) emission originates from the upscattering of optical photospheric photons into the X-ray energy range by relativistic electrons accelerated at the SN shock. When the optical SN is bright, IC emission can dominate the X-ray luminosity. The optical bolometric light-curve of SN\,2013by peaks at $\sim10$ days post-shock breakout \citep{Val15}.

Assuming that IC dominates around the optical peak, and adopting the formalism of \citet{Margutti12} modified for massive star density profiles \citep{Margutti14b}, we infer a mass-loss rate of $\dot M\sim 10^{-6}\,\rm{M_{\sun}yr^{-1}}$ for a wind velocity of v$_w=$10 km\,s$^{-1}$. This calculation assumes an electron distribution $N_e\propto \gamma^{-p}$ with power-law index $p=3$ and a post-shock energy fraction into relativistic electrons  $\epsilon_e=0.1$, which  are standard values in SN modeling (e.g.  \citealt{Chevalier06}), and $n=10$, where the ejecta density profile scales as $\rho_{sn}\propto v^{-n}$. The X-ray luminosities do not account for intrinsic absorption, which may be important at early times, and thus the inferred mass-loss rate represents a lower limit to the true value. 

The flat temporal evolution of the detected X-ray emission after optical peak suggests that other radiative processes take over at this time. First, we derive a rough upper limit on $\dot M$ by assuming that IC emission is the only source of X-rays at $\sim35$ days post explosion and obtain $\dot M\sim 10^{-4}\,\rm{M_{\sun}\,yr^{-1}}$ (v$_w=$10 km\,s$^{-1}$), shown in Figure \ref{fig:IC}.

At $t\sim70$ day, the optical emission from SN~2013by is significantly fainter and IC emission is likely negligible.  Interpreting the detected X-ray luminosity as due to free-free thermal emission from the reverse shock in a wind medium, and adopting the formalism by \cite{Chevalier06} for an oxygen dominated ejecta, we infer $\dot M\sim (2-4)\times 10^{-6}\,\rm{M_{\sun}\,yr^{-1}}$ (v$_w=$10 km\,s$^{-1}$) for a range of $n=10-15$, shown as the blue line in Figure \ref{fig:IC}. However, we caution that our calculations should be treated as order of magnitude estimates due to the sparse coverage of the X-ray data and the limited spectral information.

\subsection{The SN to SNR Transition}

Optical emission from supernovae in the first few months to years after explosion is powered by the radioactive decay chain $^{56}$Ni$\rightarrow$$^{56}$Co$\rightarrow$$^{56}$Fe. In Type II events the dominant emission line is H$\alpha$ with typical FWHM velocities of $\sim 3000$ km\,s$^{-1}$.

As the forward shock encounters increasing amounts of CSM, a reverse shock develops that propagates into outward expanding ejecta, which then becomes heated and ionized \citep{Chevalier94}. Due to the reverse shock interacting with high-velocity ejecta, these velocities are distinctly larger than those observed at earlier epochs ($\la 1$ yr after explosion). Indeed, the majority of CCSNe that have been observed more than two years after explosion have line velocities of 5000 $-$ 10,000 km~s$^{-1}$ \citep{Fesen99}.  

It is the formation of a reverse shock where the ejecta become shock heated that the SN begins to transition towards becoming a SNR \citep{Mil12,Mil17}. This transition is rarely observed. Only in a handful of nearby cases has it been possible to monitor core collapse explosions several years to decades after maximum light. The timescale of the SN-to-SNR transition is strongly dependent on the explosion dynamics, the progenitor structure, and the  properties of local CSM/ISM environment it encounters.  

In SN~2013by, we appear to be observing the SN-SNR transition within the first two years after the explosion.  Specifically, we interpret the large Doppler broadening of H$\alpha$ emission line profile observed in SN\,2013by one year post-max to be due to outer H-rich ejecta being excited by the passage of the reverse shock.

In Figure \ref{fig:ReverseShock}, we highlight this transition in SN\,2013by and compare it to examples observed in SNe IIP. Emission observed during epochs prior to  broadening are associated with heating by radioactive $^{56}$Co (blue), whereas those where broadening is observed are associated with passage of the reverse shock through outer H-rich ejecta (red), where the reverse shock spectra are shaded to emphasize the evolution between the early and late epochs.

Comparing SN~2013by to two Type IIP SNe, SN\,2013ej and SN\,2004et, the broad H$\alpha$ emission in SN\,2013by seen at day +457 associated with reverse shock interaction is comparatively weak.  The spectrum of SN~2013ej on day +807 is dominated by its broad, flat-topped H$\alpha$ profile that extends out to velocities of roughly -9000 and +10,000 km~s$^{-1}$ \citep{Mau16}. Such flat-topped or `square' emission line profiles can be produced by a spherical shell of material that becomes excited by a reverse shock.

Although the H$\alpha$ profile seen in SN\,2004et extends out to comparable velocities of SN\,2013ej, it is not completely flat-topped.  This may be due to components from both $^{56}$Co heating and reverse shock interaction. The central, lower velocity (FWHM <~3500 km~s$^{-1}$ at +278 days) component rises above the higher velocity ($\sim$7000 km~s$^{-1}$) flat-topped component. The broad H$\alpha$ profile also exhibits the broad blue-red asymmetry seen in SN\,2013ej.  Mid-infrared observations support the notion that the asymmetry seen here is due to the formation of dust in the ejecta \citep{Kot09}.

In SN~2013by, the broadening seen in the blue, increasing from -4000 km\,s$^{-1}$ on day +278 to at least -7500 km\,s$^{-1}$ on day +457.  Interestingly, the broadening of H$\alpha$ in SN\,2013by is apparent at a significantly earlier epoch than the other two SNe.  While evidence of reverse shock emission in SNe~2004et and 2013ej is not visible until +800 days, SN\,2013by shows indications of this phenomenon by +457 days. However, broad [\ion{O}{1}] $\lambda\lambda$6300,6364 emission is present in both 2013ej and 2004et, but not in SN~2013by. Presumably this means that its reverse shock has not reached 2013by's O-rich ejecta at this epoch.

\subsection{The SN~2013by Progenitor}\label{ProgSystems}

Unlike SNe IIP, SNe IIL may belong to a class of progenitor systems where the effects of binary evolution strongly affect the outcome of the observed supernova explosion and may be the first in an evolutionary link of increasingly stripped stars between IIL $\Rightarrow$ IIb $\Rightarrow$ Ib SNe \citep{Nom96}.  Analysis of the SN-to-SNR transition can provide clues about the progenitor star's evolution, including its possible binary nature and mass loss in the poorly understood phases approaching core collapse. 

Type IIL supernovae are thought to be associated with stars that have relatively low mass ($\approx 1-2$~M$_{\odot}$) and/or extended hydrogen-rich envelopes with radii as large as regular red supergiants ($\sim$200~R$_{\odot}$). These low-density envelopes allow the shock break out velocity at the bottom of the H-rich envelope to be higher, thus sending a stronger blastwave into surrounding ISM/CSM compared to their Type IIP counterparts that have much more massive and dense envelopes. 

This framework is consistent with the three objects we have considered here.  In the SNe~IIP objects, SN\,2013ej and SN\,2004et, the Doppler broadening was observed at epochs twice as late as that observed for SN\,2013by.  Hence, an early transition to the reverse shock phase may be a characteristic of SNe IIL. It may also explain why many SNe IIL such as SN\,1970G, SN\,1979C, and SN\,1980K are visible decades after explosion \citep{Fesen93,Fesen99,Mil09}.

Moreover, analysis of the X-ray light curve provides insight into the progenitor's environment prior to explosion.  The flat temporal evolution of the X-ray light curve suggests that there is an additional flux contribution from a source other than IC emission at later times.  Around day +70 free-free emission due to the reverse shock interacting with the wind medium dominates.

Based on the X-ray models shown in Figure~\ref{fig:IC}, SN\,2013by likely exploded in a fairly dense local medium, enriched by sustained mass-loss from the stellar progenitor with $\dot M\sim 10^{-4}-10^{-6}\,\rm{M_{\sun}yr^{-1}}$. This is well within the range of mass loss rates estimates for progenitor systems of other SNe IIL \citep{Weiler02}.


\section{Conclusions}

We have presented X-ray, optical, and near-infrared observations of the Type IIL SN~2013by.  These multi-epoch observations provide a look into the late time evolution of a SN IIL and allow us to infer the properties of the progenitor star and mass loss history.  We extend the light curve of SN\,2013by from \citet{Val15} out to late times to create a light curve that spans from day -8 to day +457.  Our main results and conclusions are as follows:

1) Spectral and photometric properties indicate that SN\,2013by quickly interacted with local CSM.
 For early data up to day +278, modeling of SN\,2013by's optical and X-ray light curves requires energy sources in addition to radioactive $^{56}$Co heating and IC upscattering, respectively. SN-CSM interaction is the most natural explanation. The triple-peaked H$\alpha$ emission line profile of SN\,2013by observed on days +151 and +278 is not unlike that observed in SN\,1998S, which was interpreted to be the consequence of H-rich ejecta interacting with nearby dense CSM distributed in a disk along the equatorial plane.  

2) The SN-SNR transition time frame in SN\,2013by (between days +278 and +457) is faster than that observed in these other CCSNe objects at much later times, suggesting that reverse shock formation may occur earlier for SNe IIL than for SNe IIP. This is consistent with the notion that SNe IIL have less massive and/or more extended H-rich envelopes due to mass loss as compared to those of SNe IIP at the time of explosion.

3) On day +457, SN~2013by's H$\alpha$ emission feature exhibited significant broadening from roughly $-$4000 km~s$^{-1}$ at day +278 to $-$7500 km~s$^{-1}$, which is consistent with the emergence of a reverse shock exciting high velocity H-rich ejecta.  The observed flat light curve of the X-ray light curve is indicative of CSM interaction and the formation of the reverse shock.  We suggest that SN~2013by's progenitor underwent sustained mass-loss $\sim$ $10^{-6}\,\rm{M_{\sun}yr^{-1}}$ just prior to explosion.

4) We detect the first overtone of CO at 2.3 $\micron$ in the NIR spectrum at day +152.  This is the first time CO has been detected in a Type IIL SN.  The presence of CO is indicative of the early formation of dust and is consistent with the blue-red asymmetry observed in the emission line profiles (Figs. \ref{fig:Ha} and \ref{fig:all_vel}).

\vspace{0.5cm}
Models of SN-CSM interaction predict how the emission will evolve with time as the reverse shock penetrates into deeper layers of ejecta \citep{CF16}. It is anticipated that the velocity widths of emission lines in SN~2013by will narrow with time and that the relative line strengths should change as expansion drives down ejecta density. Specifically, the flux ratio [O~III]/([O~I]+[O~II]) should increase with time, and H$\alpha$/([O~I]+[O~II]) should decrease \citep{Mil12}.

Additional late-time observations of other SNe~IIL may help clarify the speed of the SN-SNR transition in these objects.  The timescale of the SN-SNR transition provides useful information about the mass and distribution of the progenitor star's mass loss in the terminal stages leading to core collapse.  Establishing time lines for the SN-SNR transition relative to Type IIL and IIP SNe will lead to a better understanding of the nature of the CSM environment around Type II SNe and build a more complete picture of their progenitors.

\acknowledgments
The authors wish to thank J. Mauerhan for sharing his late-time spectra of SN\,2013ej and S. Valenti for sharing his photometry of SN\,2013by. C.S.B.'s research is supported in part by a Fellowship from Dartmouth's School of Graduate and Advanced Studies. This paper includes data gathered with the 6.5 meter Magellan Telescopes located at Las Campanas Observatory, Chile.

\bibliographystyle{apj}
\bibliography{Bib_master.bib}

\end{document}